\documentclass[a4paper,11pt]{article}
\usepackage{jcappub}
\usepackage{lineno}
\usepackage{amsmath}
\usepackage{algorithm}
\usepackage{algorithmicx}
\usepackage{algpseudocode}

\arxivnumber{2507.09956}
\title{\boldmath Constraining $\sigma_8$ with Lensing Statistics in Low and High Density Regions} 

\author[a]{Yiqi Huang,}
\affiliation[a]{State Key Laboratory of Dark Matter Physics, School of Physics and Astronomy, Shanghai Jiao Tong University, Shanghai 200240, China}

\author[b,c]{Fuyu Dong,}
\emailAdd{dfy@ynu.edu.cn}
\affiliation[b]{South-Western Institute for Astronomy Research, Yunnan University, Kunming, Yunnan 650500, China}
\affiliation[c]{Key Laboratory of Survey Science of Yunnan Province, Yunnan University, Kunming, Yunnan 650500, China}

\author[a]{Jun Zhang,}
\emailAdd{betajzhang@sjtu.edu.cn}

\author[a]{Cong Liu}

\author[d]{and Hekun Li}
\affiliation[d]{Shanghai Astronomical Observatory, Chinese Academy of Sciences, 
Shanghai 200030, China}

\abstract
{Lensing studies are typically carried out around high density regions, such as groups and clusters, where the lensing signals are significant and indicative of rich density structures. However, a more comprehensive test of the cosmological model should also include the lensing effect in low density regions. 
In this work, we incorporate the stacked weak lensing signals around the low density positions, alongside galaxy-galaxy lensing and galaxy-galaxy two point correlation function to perform a joint cosmological analysis on $\sigma_8$. The low density positions are constructed from the DR9 data release of the DESI legacy imaging survey, using galaxies with r-band absolute magnitude cut M$<$-21.5 and in the redshift range of 0.18$<$z$<$0.28. In doing so, we simultaneously parameterize photometric redshift errors and halo mass uncertainties while building mock catalogs from simulations using the method of SubHalo Abundance Matching (SHAM). For the weak lensing measurements, we use the shear estimators derived from the DECaLS DR8 imaging data, processed by the Fourier\_Quad pipeline. The survey boundaries and masks are fully taken into account. Our analysis achieves a total significance of $31.1\sigma$ detection for lensing in the low density positions, which significantly improve the $\sigma_8$ constraint compared to galaxy-galaxy lensing and galaxy-galaxy two point correlation function by 14$\%$. For flat $\Lambda$CDM model, we constrain $\sigma_8$ =$0.824^{+0.015}_{-0.015}$, which shows a good agreement with the PLANCK result. Additionally, the halo mass uncertainty $\sigma_{\lg M}$ and photometric redshift error $\sigma_z$ are constrained to be $0.565^{+0.086}_{-0.070}$ and $0.004^{+0.004}_{-0.003}$ respectively, which are somewhat different from our expectations due to the significant degeneracy of the two parameters.}

\begin{document}
\maketitle
\flushbottom
\section{Introduction} \label{sec:intro}

Weak lensing refers to the small but coherent distortions of the background galaxy shapes by the foreground density fluctuations \cite{2001PhR...340..291B}. 
As the statistics of such background distortions are directly determined by the properties of the foreground density field, weak lensing has become a popular and powerful probe of the large scale structure in recent galaxy surveys, including Canada-France-Hawaii Telescope Lensing Survey\footnote{\url{www.cfhtlens.org}}(CFHTLenS), Dark Energy Survey\footnote{\url{www.darkenergysurvey.org}}(DES), Hyper Suprime-Cam Subaru Strategic Program\footnote{\url{hsc.mtk.nao.ac.jp/ssp}}(HSC), and Kilo-Degree Survey\footnote{\url{kids.strw.leidenuniv.nl}}(KiDS). With the development of the stage IV surveys, like Elucid\footnote{\url{www.euclid-ec.org}}, Large Synoptic Survey Telescope\footnote{\url{www.lsst.org}}(LSST), China Space Station Telescope\footnote{\url{www.nao.cas.cn/csst}}(CSST), there exists a great potential for weak lensing in enhancing the constraining power for the cosmological parameters.

There has been a mature development of weak lensing statistics, such as galaxy-galaxy lensing \cite{1996AAS...188.1302B, 2013MNRAS.432.1046B, 2017MNRAS.465.4204C, 1998ASPC..136..323H, 2018PhRvD..98d2005P, 2002sgdh.conf...58W, 2011A&A...534A..14V}, 
shear-shear correlation \cite{2007PhRvD..75d3010F,2015RPPh...78h6901K,2018PhRvD..98d3528T,2019PASJ...71...43H,2021A&A...645A.104A}, peak counts \cite{2000ApJ...530L...1J,2010MNRAS.402.1049D,2010PhRvD..81d3519K,2015PhRvD..91f3507L}, mass map making \cite{2015PhRvD..92b2006V, 2018MNRAS.475.3165C}, Minkovski functionals \cite{2012PhRvD..85j3513K,2012MNRAS.419..536M,2014ApJ...786...43S,2015PhRvD..91j3511P}, and also those high-order statistics such as the three-point correlation function \cite{2003ApJ...583L..49T, 2003MNRAS.340..580T}; The probability distribution function \cite{2022PhRvD.105d3537B, 2016PASP..128j4502S}; The mass aperture moment \cite{2022PhRvD.105j3537S, 2022PhRvD.106h3509G}; And the integrated shear three-point correlation function \cite{2023AAS...24124405H, 2023JCAP...07..040G}.
Such investigations are usually applied to the high density regions to study the properties of halos \cite{2018ApJ...862....4L} or to constrain cosmological parameters \cite{2009MNRAS.394..929C}.
While in regions of low densities, such as cosmic voids \cite{2009MNRAS.400.1835B, 2010MNRAS.403.1392L, 2014MNRAS.443.2983S}, we expect nonlinear evolutions and baryonic effects to be minor.
Consequently, by integrating measurements from low density fields, it becomes possible to enhance the constraining power of cosmological parameters such as $\Omega_m$ and $\sigma_8$. And as for the explanation of cosmic acceleration, weak lensing is more distinguishable in low density fields between dark energy and modified gravity models \cite{2015MNRAS.450.3319L}.

More recently, there have been proposals that the lensing signals around galaxies from observations are systematically lower than the predictions from mocks using N-body simulations \cite{2017MNRAS.467.3024L} (though see \cite{2024arXiv240704795C} and \cite{2025arXiv250209404L} for counter arguments). They measure galaxy-galaxy lensing for the Baryon Oscillation Spectroscopic Survey (BOSS)\citep{2011AJ....142...72E} CMASS sample using weak lensing data from CFHTLenS and the Canada France Hawaii Telescope Stripe 82 Survey (CS82)\citep{2014RMxAC..44..202M} of about 250 square degrees and compare with predictions from mock catalogs generated using the clustering of the CMASS galaxies and the standard galaxy-halo connection model. The results from the mock are systematically higher than the observed by 20-40$\%$. There also emerged some studies based on similar data and model\cite{2020MNRAS.492.2872W,2020MNRAS.491...51S,2021MNRAS.502.2074L}. Meanwhile, recent measurements of the shear-shear correlations from several different weak lensing surveys also indicate that the $S_8$ value from weak lensing are systematically lower than that from Planck \cite{2021A&A...646A.140H,2022PhRvD.105b3520A}. 
But recently \cite{2025arXiv250319441W} claims that after the improvement of redshift distribution estimation and calibration, the updated parameter $S_8=0.815^{+0.016}_{-0.021}$ is consistent (0.73$\sigma$) with PLANCK. Note that the measurements on the shear-shear correlation suffer from systematic issues (e.g., source redshift distribution, correlated shear bias) that are very different from those associated with galaxy-galaxy lensing or galaxy clustering (e.g., galaxy selection due to observational effects, such as fiber collision \cite{10.1093/mnras/stx185}).
In this situation, we believe it is timely to study the lensing effect in the low density regions to further check the consistency of the $\Lambda$CDM model.


Identification of voids is typically carried out in 3D using galaxies' spectroscopic redshift information\cite{2002ApJ...566..641H,2005MNRAS.360..216C,2013MNRAS.434.2167J,2015MNRAS.448..642E}. However, spectroscopic surveys are expensive, and often contain significant selection effects.
Alternatively, one can define the low density regions using the projected galaxy distributions \cite{2015MNRAS.450.3319L,2016MNRAS.456.2662F, 2016MNRAS.455.3367G}, such as the density split statistics \cite{2018PhRvD..98b3507G}, which only requires photo-z information. They divide the sky into cells and assign each a weighted and smoothed galaxy count, then measure the shear signals for cells of different galaxy counts. \cite{2019ApJ...874....7D} proposes a similar but different approach: it defines the so-called low density positions (LDPs) as those that are away from the foreground bright galaxies by more than a critical radius in projection. This definition avoids the 'void center' problem which exists in most other void finder algorithms, on the other hand, the stacked lensing signals around LDPs generally yield a much higher significance due to the large number of LDPs; Compared to the density split statistics, it is easier to achieve without calculating the galaxy density map, and could more easily build up the relation between galaxies and (sub)halos in simulations. Within this framework, \cite{2021MNRAS.500.3838D} successfully measured the integrated Sachs-Wolfe effect by cross-correlating the CMB temperature map with the LDPs. \cite{2021ApJ...923..153D} detected the cross-correlation between CMB Lensing and LDPs, further demonstrating that the LDPs are very useful tracers of the low density regions.

The main goal of this paper is to perform a joint analysis of LDP lensing, galaxy-galaxy lensing (GGL) and the galaxy two-point correlation function (2PCF) to constrain the cosmological parameter $\sigma_8$ (by fixing $\Omega_m$ and other cosmological parameters). As shown later in the paper, a direct comparison between the observational results and the simulation ones have quite significant discrepancy ($\sim 20-30\%$), which we believe is largely due to the photometric redshift errors ($\sigma_z$) and the uncertainties/scatters in the halo mass - galaxy luminosity correspondence ($\sigma_{\lg M}$). We therefore try to consider these errors in our analysis with the help of the N-body simulations.

This paper is organized as follows: In \S\ref{galaxy sample}, we introduce our lens and source catalogs. \S\ref{statistics} contains the statistical method and the details of the measurements. In \S\ref{simulation} we describe our procedures in simulations for reproducing the observational results. We show our main results in \S\ref{result}, including the constraints on $\sigma_8$, $\sigma_z$, and $\sigma_{\lg M}$. In \S\ref{sec:method validation}, we test our method using simulations. We summarize our findings and discuss related issues in the last section. In our study, for the cosmological parameters other than $\sigma_8$, we use the PLANCK results \citep{PLANCK2018}: $\Omega_m$=0.315, $\Omega_\Lambda$=0.685, $h$=0.673.

\section{Galaxy \& Shear Catalog}
\label{galaxy sample}\label{shear catalog}

Our study is based on the galaxy catalog from the Dark Energy Camera Legacy Survey (DECaLS), which is one of the Dark Energy Spectroscopic Instrument (DESI) Legacy Imaging Survey \cite{2019AJ....157..168D} programs.
DESI covers around 14000 deg$^2$ in Northern and Southern Galactic caps in three optical bands ($g,r$ and $z$), and the median 5$\sigma$ point source depths reach 24.9, 24.2 and 23.3 respectively.
The photometric redshift catalog is from \cite{2021MNRAS.501.3309Z}, who uses a machine learning algorithm based on decision trees, i.e., the random forest regression method, to determine the photo-z. The training data is from ten different surveys that have overlap with the DECaLS footprint. 

The mask used in this study is provided on the official DECaLS website \footnote{\url{http://legacysurvey.org/dr9/files/*random-catalogs}},  
which is well distributed and can produce the same sky coverage with the galaxy population. It provides the exposure times for g, r, z bands for each random point. We select those with nonzero exposure time in all three bands and the item 'MASKBITS' of this catalog is equal to zero as our random points. For each 0.5$\times$0.5 arcmin$^2$ grid (considering the galaxy number density and the radius we set to determine the LDPs), we select the center of each with no random points to be our mask points.

The shear catalog of the background sources used in this study is also based on the photometric catalog of the DESI Legacy Imaging Surveys from DR8.
Our shear catalog is generated by the Fourier$\_$Quad (FQ hereafter) pipeline \cite{2008MNRAS.383..113Z,2015JCAP...01..024Z,2017ApJ...834....8Z,2022AJ....164..128Z}.
There are five shear estimators in the FQ method
$G_1, G_2, N, U, V$, which are derived from the multiple moments of the galaxy image power spectrum in Fourier space:
\begin{eqnarray}
\begin{aligned}
G_1&= -\frac{1}{2}\int d^2\vec {k}(k_x^2-k_y^2)T(\vec {k})M(\vec {k})\\
G_2&=-\int d^2\vec {k} k_x k_y T(\vec {k})M(\vec {k})\\
N&=\int d^2\vec {k} [k^2-\frac{\beta^2}{2}k^4] T(\vec {k})M(\vec {k})\\
U&=-\frac{\beta^2}{2}\int d^2\vec {k} (k_x^4-6k_x^2k_y^2+k_y^4) T(\vec {k})M(\vec {k})\label{U}
\\
V&=-2\beta^2\int d^2\vec {k} (k_x^3k_y-k_xk_y^3) T(\vec {k})M(\vec {k})
\end{aligned}
\end{eqnarray}
where $k$ is the wave vector, $M(\vec {k})$ is the galaxy power spectrum corrected by subtracting the contributions from the background noise and the Poisson noise. $T(\vec {k})$ is used to transform the Point Spread Function (PSF) to an isotropic Gaussian form. They are defined as:
\begin{eqnarray}
\begin{aligned}
& M(\vec {k})=|\tilde{f}^S(\vec {k})|^2-\tilde{F}^S-|\tilde{f}^B(\vec {k})|^2+\tilde{F}^B \\
&T(\vec {k})=|\tilde {W}_{\beta}(\vec {k})|^2/|\tilde{W}_{PSF}(\vec {k})|^2 
\end{aligned}
\end{eqnarray}
where
\begin{equation}
\tilde{W}_{\beta}(\vec {k})=\frac{1}{2\pi \beta ^2}\exp(-\frac{|\vec{k}|^2}{2\beta ^2}),    
\end{equation}
is the Gaussian kernel with scale radius $\beta$, which should be somewhat larger than that of the original PSF \cite{2015JCAP...01..024Z}. $f$ is the galaxy surface brightness, $\tilde{f}^S(\vec {k})$ and $\tilde{f}^B(\vec {k})$ denote the Fourier transform term of the source galaxy and the background, $\tilde{F}^S$ and $\tilde{F}^B$ are the power spectrum of the Poisson noise estimation. One can show that the ensemble averages of the shear estimators can recover the underlying shear signals to the second order in accuracy:
\begin{equation}
\frac{\langle G_i\rangle}{\langle N\rangle}=g_i+O(g_{1,2}^3) (i=1,2). 
\end{equation}

In practice, instead of taking the ensemble averages, we use the so-called PDF-SYM method to derive the underlying shear signals or their various statistics from the FQ shear estimators \cite{2017ApJ...834....8Z}. The idea is to find the best assumed value of the shear component $g_{1/2}$ that can symmetrize the probability distribution functions (PDF) of the shear estimator $G_{1/2}-g_{1/2}(N\pm U)$ around zero. It does not require weightings on the shear estimators, and the statistical uncertainties on the results could automatically approach the Cramer-Rao bound (the lower bound of the statistical uncertainty).

The FQ shear catalog was tested directly using the field distortion (FD hereafter) effect (optical aberration) on the focal plane \cite{2019ApJ...875...48Z}. By comparing the FD signals derived from the astrometric calibration with those from the shear estimators of galaxies, one can get the multiplicative and additive biases of the real data explicitly. In this work, we use the z-band shear catalog, which shows negligible biases according to the FD test \cite{2022AJ....164..128Z}. This shear catalog has been successfully applied in a number of recent works to study the dark matter halo properties and cosmological parameters \cite{2022ApJ...936..161W,Ziwen_2022,2022MNRAS.513.4754F,2023ApJ...947...19A,Xu_2023,2024SCPMA..6770413L,Zhang_2024}.

\section{Statistical method}\label{statistics}
\begin{figure}
    \centering
    \includegraphics[width=0.65\textwidth]{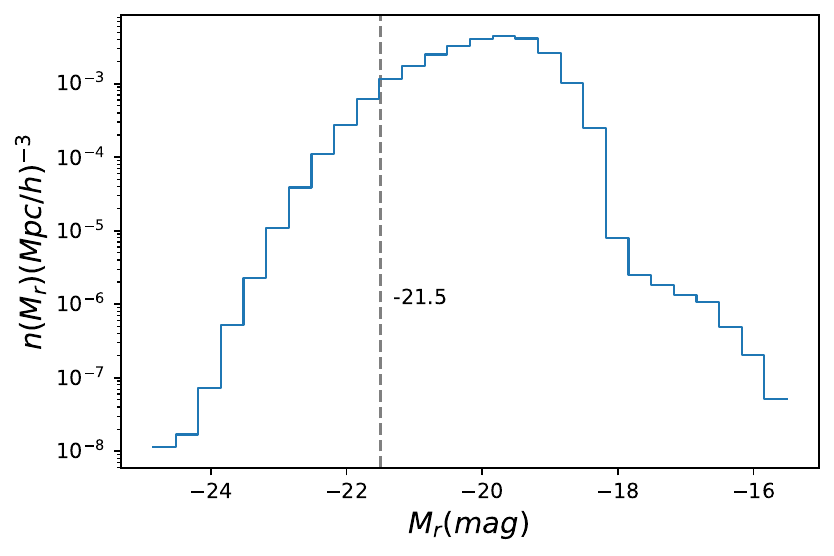}
    \caption{\label{Fig:pdf_obs}
    {The absolute magnitude distribution of galaxies in the redshift range between 0.18 and 0.28. The
grey dotted line refers to the cut at -21.5 for FBGs. 
}}
\end{figure}

\begin{figure}
    \centering
    \includegraphics[width=0.65\textwidth]{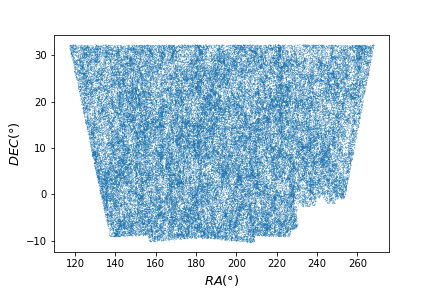}
    \caption{\label{Fig:footprint}
    {Footprint of the DECaLS galaxy samples within the redshift range between 0.18 and 0.28, and with absolute magnitude smaller than -21.5. 
}}
\end{figure}

In this study, we choose a continuous sky area of about 4800 deg$^2$ from the North Galactic Cap (DEC$\le 32^{\circ}$) of DECaLS DR9\footnote{https://gax.sjtu.edu.cn/data/DESI.html}, 
considering the limitation of our simulation box and the continuity of the sky area. We did not include the Southern Galactic Cap nor those small isolated area in the North. 
For the purpose of defining the low density regions (LDPs) later in the paper, we choose the foreground bright galaxies (FBGs) according to their photometric redshift and the r-band absolute magnitude, the later of which is determined via:
\begin{equation}
    M-5\lg h=m-k_{correct} - 5\lg(\frac{D_L}{h^{-1}{\mathrm Mpc}})-25,
    \label{equ1}
\end{equation}
where $D_L$ is the luminosity distance determined by photometric redshift. We also take the k-correction into account. We choose the r-band absolute magnitude cut to be -21.5 and redshift from 0.18 to 0.28 throughout this paper to guarantee the completeness of the sample. The distribution of the galaxy luminosity is shown in Fig.\ref{Fig:pdf_obs}. In order to get the same footprint later in simulation, we manually cut the left and right side for a regular border. Fig.\ref{Fig:footprint} shows the footprint of the galaxy samples.  

Based on the positions of the FBGs, we can determine the locations of the LDPs by following the procedures first introduced in \cite{2019ApJ...874....7D}.
For the convenience in the rest of the discussions, for any point in the sky area of our interest, let us define $R_s$ as the shortest projected distance that this point is away from any FBGs. In \cite{2019ApJ...874....7D}, the LDPs are simply defined as those with $R_s$ larger than a critical value, say $1 ^{\prime}$. In this paper, instead, for a given set of FBGs, we divide the LDPs into three groups: I. $2^{\prime} < R_s \le 5^{\prime}$; II. $5^{\prime} < R_s \le 7^{\prime}$; III. $ R_s > 7^{\prime}$.

\begin{figure}
    \centering    \includegraphics[width=0.65\textwidth]{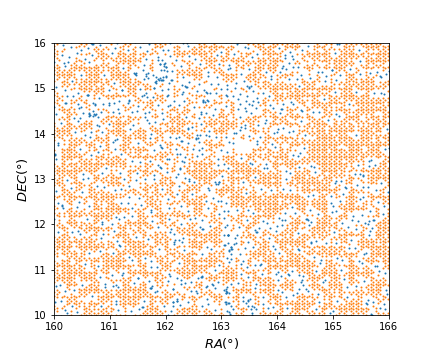}
    \caption{\label{Fig:ldp_dis_obs}
    {The distribution of the DECaLS galaxy sample (blue points, M$<$-21.5, $0.18 < z < 0.28$) and the LDPs (orange points) with the critical radius $R_s$ being $5^{\prime}$. The white areas are either masks or the neighborhood of galaxies.
}}
\end{figure}

Our LDPs are placed on a grid generated by HEALPix\cite{2005ApJ...622..759G} with $N_{side}=4096$, corresponding to the angular resolution of $0.859^{\prime}$. 
To avoid the influence of the masks, we exclude those LDPs whose distance to the mask points are less than 0.5 arcmin\footnote{We have also chosen 1 arcmin in a test, and we find that this choice does not change our main conclusion much.}. To avoid potential boundary effects, 
we require the LDPs to be 1.5$^{\circ}$ away from the boundary of the FBGs. For the irregular border in the lower part of Fig.\ref{Fig:footprint}, we cut at $DEC=-7^{\circ}$ for
$RA\le222^{\circ}$, and $DEC=2^{\circ}$ for $RA> 222^{\circ}$. 
As an example, Fig.\ref{Fig:ldp_dis_obs} shows a small fraction of our data. The blue dots represent the FBGs cut at the absolute magnitude of -21.5. The orange points are the LDP points with $R_s$ larger than 5 arcmin, and the white areas are the neighborhood of either the FBGs or the masks.

Given the FBGs and the LDPs in three groups, we consider three types of statistics: 

1. 2PCF of the FBGs, i.e. their spatial clustering in projection; 

2. the stacked excess surface density (ESD) profile of the FBGs (i.e. GGL); 

3. the stacked ESD profile of LDPs. 

The 2PCF of the FBGs is measured by calculating the Hamilton Estimator \cite{1993ApJ...417...19H} for different radius bins:
\begin{equation}
     \xi(r)=\frac{DD(r)\times RR(r)}{DR(r)^2}-1
\end{equation}
Here $r$ represents the pair separation perpendicular to the LoS in comoving units, which is indeed transformed from the angular separation assuming both galaxies are at redshift 0.23. $DD(r)$ represents the pair number of FBGs at distance $r$, and $RR(r)$ is the number of random pairs. $DR(r)$ is the number of pairs including one FBG and one random point. The random points cover the same footprint as the FBGs, and ten times more.
The random points used here are not from the official catalog, but generated separately by excluding the masked area.

In the ESD measurements, we use the relation between the ESD and the tangential component of the shear signal $\gamma_t$:
\begin{equation}
    \Delta\Sigma(R) = \gamma_t\Sigma_{c}=\bar\Sigma(\leq R)-\Sigma(R)
    \label{deltasigma}
\end{equation}
where $\bar\Sigma(\leq R)$ is the averaged surface density within radius $R$, $\Sigma(R)$ is the surface density at radius $R$, and $\Sigma_c$ is the critical surface density defined in comoving unit:
\begin{equation}
    \Sigma_c=\frac{c^2}{4\pi G}\frac{D_s}{D_lD_{ls}(1+z_l)^2}
\end{equation}
where G is gravitational constant, c is the speed of light, $z_l$ is the redshift of the lens, $D_s, D_l$ and $D_{ls}$ are the angular-diameter distances between the observer and the source galaxies, the observer and the lens, the lens and the source respectively.

Our ESD measurements are carried out using the PDF\_SYM method, in which the ESD is estimated by symmetrizing the PDF of the modified tangential shear estimator using the assumed value of the ESD $\widehat{\Delta\Sigma}$:
\begin{equation}
    G_t-\frac{\widehat{\Delta\Sigma}}{\Sigma_c}(N+U^{\prime})
\end{equation}
where $G_t$ is the tangential component of the FQ shear estimator, and $U^{\prime}$ is the corresponding term of $U$ after coordinate rotation.

In our measurement, we are aware that additional shear bias may be caused by the redshift cut or the weight of $\Sigma_c$, we therefore need to perform an {\it onsite} calibration of the shear bias using the field distortion signal, as what is done in \cite{2025JCAP...01..068S}. This is a special advantage of the FQ shear catalog, as each of its shear estimators is measured on a single exposure, and the field-distortion information is kept in the catalog. The details of this operation for this work is shown in the Appendix.

\section{Simulation} 
\label{simulation}

The theoretical predictions of the statistics mentioned in the previous section are made through numerical simulations. The CosmicGrowth (CG hereafter) Simulations contain a suite of high resolution N-body simulations with different cosmologies \cite{2002ApJ...574..538J,2019SCPMA..6219511J}. We use the one corresponding to Planck cosmology in this work. The name of the simulation is 'Planck\_2048\_1200', referring to the box size of 1200$Mpc/h$, and the particle number of 2048$^3$. 
Halos and subhalos are found using FoF and the Hierarchical Branch Tracing (HBT) \cite{2012MNRAS.427.2437H,2018MNRAS.474..604H} algorithms. Totally we get 24 snapshots with even logarithmic intervals
in the scale factor from redshift 72 to 0.

\subsection{Lightcone Construction}\label{lightcone}
To build the lightcone, we take nine snapshots of CG with redshift from 0.512 to 0 (z=0.512, 0.435, 0.363, 0.295, 0.23, 0.168, 0.11, 0.05, 0). The observer is assumed to be standing at the center on one side of the simulation box. Because of the large sky area of DECaLS' footprint, we add two boxes at the left and right sides using the periodic boundary condition.  For each (sub)halo and dark matter particle in each snapshot, we transform the Cartesian coordinates to the corresponding right ascension (RA), declination (Dec), and comoving distance to the observer. The lightcone is then formed by stacking the spherical shells of (sub)halos and particles from different redshift snapshots according to their distances to the observer.

The redshift range for the FBGs we consider is from 0.18 to 0.28. Taking into account the redshift uncertainty ($\sim 0.03$, see detail in \S\ref{4.4}), the major structure in our measurement occupies about 400$Mpc/h$ along LoS, occupying the central 1/3 portion of the whole box. To achieve a more precise measurement, we also put the observer at 400$Mpc/h$ and 800$Mpc/h$ away on the LoS to use the front and back 1/3 portion of the box. Note that to do so, we also need to extend the box volumn using the periodic boundary conditions. Note that the three portions along the LoS would contain minor overlaps. Given that there are three orthogonal axes, we repeat the measurement along the other two axes as well. In total, we would get nine different lightcones. We perform our calculation for these lightcones through the same procedure, and use their average results as our theoretical prediction.
  
\subsection{Linking (Sub)Halos to Galaxies}\label{sham}
In order to mimic the observation, we use the SubHalo Abundance Matching (SHAM)\cite{2004MNRAS.353..189V, 2006ApJ...647..201C, 2008MNRAS.388..945B} method to link (sub)halos to galaxies. After the lightcone has been constructed, we choose (sub)halos within the same redshift range as in observation. Besides, we also add the mask of DECaLS to the simulation.
Under the assumption that massive halos contain brighter galaxies, we construct the connection between (sub)halos and the galaxies by counting their numbers
\begin{equation}
    \label{equ:sham}
    \int_L^{\infty}\phi (L)dL=\int_M^{\infty}[n_h(M)+n_{sh}(M)]dM
\end{equation}
where
\begin{align*}
\begin{split}
M= \left \{
\begin{array}{ll}
    M_{acc},    & subhalos\\
    M_z,     & distinct\, halos,
\end{array}
\right.
\end{split}
\end{align*}
where $\phi(L)$ is the luminosity function, $n_h/n_{sh}$ is the number of halos/subhalos, and $M_{acc}$ is the accretion mass of a halo (mass before it mergers to another halo), which is commonly used in the SHAM models for the relation to the halo merger history. $M_z$ is the halo mass at redshift z. For both halos and subhalos, we take their virial masses as 'M' in Eq.\ref{equ:sham}.  

After the (sub)halos are linked to the FBGs, we could derive the location of LDPs according to the procedures adopted in the observation. We also take into account the observational effects, such as masks, photo-z error, and the uncertainty of halo mass. For the calculation of the lensing signal, we stack the density profiles around each (sub)halo/LDP, and calculate the ESD according to Eq.\ref{deltasigma}.

\subsection{Adding Mass \& Photo-z Uncertainties in the Modeling}\label{4.4}

\begin{figure}
\centering
\includegraphics[width=0.65\textwidth]{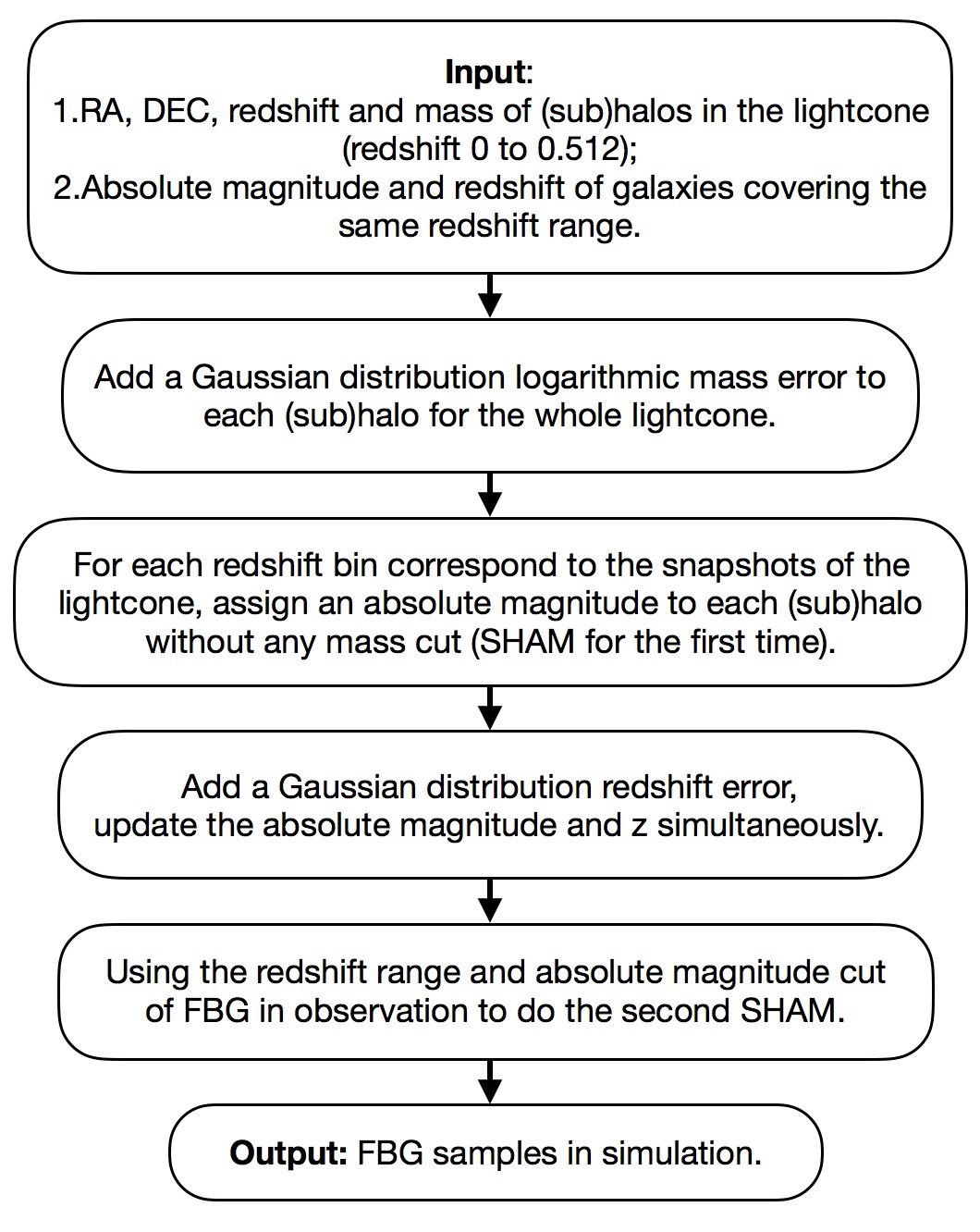}

    \caption{\label{Fig:flow chart}
Flow chart of selecting FBGs in simulation. 
}
\label{flowchart}
\end{figure}

The halo-galaxy correspondence from SHAM, as well as the photometric redshift itself, contain errors that need to be taken into account in the modeling of the statistics. We adopt a simple recipe in this paper to model the uncertainties of the halo mass and redshift. Basically, we treat the uncertainty of the common logarithm of the halo mass as a Gaussian random variable with standard deviation of $\sigma_{\lg M}$. The redshift uncertainty also obeys the Gaussian distribution with standard deviation of $(1+z)\sigma_z$, where $\sigma_z$ is from the normalized median absolute deviation defined in \cite{2013ApJ...775...93D}, which is believed to be robust in the presence of outliers.
For simplicity, we do not consider the mass dependence of $\sigma_{\lg M}$, nor the redshift dependence of $\sigma_z$ or any sort of catastrophic outliers in the redshift errors. These elements will be considered in a future work. As a result, the uncertainties are only parameterized by two quantities: $\sigma_{\lg M}$ \& $\sigma_z$. These uncertainties are all applied to (sub)halos in the simulation to mimic what we observe from the galaxies. This is done according to the procedures shown in the flowchart of fig.\ref{flowchart}. We give some more details below:

1. Before doing the first SHAM, we add a mass error to every (sub)halo in the whole lightcone from redshift 0 to 0.512, not only to the massive ones within a restricted redshift range. This is necessary as the mass errors can directly affect the mass order of (sub)halos, and thus their selection as the mock FBGs in SHAM. 

2. The first SHAM is carried out for the whole (sub)halo mass in each redshift bin (corresponding to the snapshots with redshift 0.512,
0.435, 0.363, 0.295, 0.23, 0.168, 0.11, 0.05, 0), match to the galaxies of DECaLS using the information of the r-band apparent magnitude down to 24. Each (sub)halo is assigned an absolute magnitude. We perform SHAM in different redshift bins since the halo mass function does evolve with time. Finally, a random redshift error is assigned to each (sub)halo, and its absolute magnitude is updated accordingly.

3. After both the mass and redshift errors are added, the second SHAM is performed according to the required redshift range 0.18 to 0.28 and the magnitude cut -21.5 to select out the mock FBGs in the simulation.


\begin{figure}
    \centering
    \includegraphics[width=0.65\textwidth]{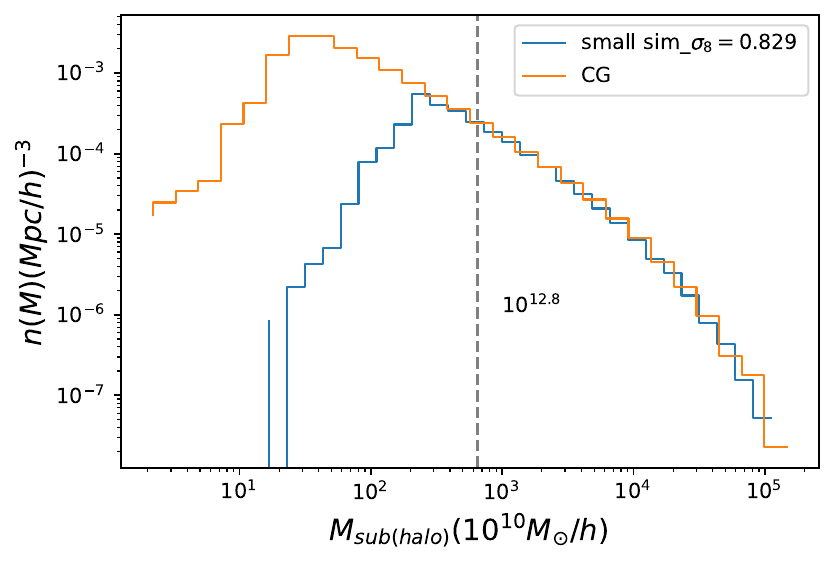}
    \caption{\label{Fig:pdf_sim}
    The (sub)halo mass
function from redshift 0.18 to
0.28 in CG (orange) and the small simulation with $\sigma_8$=0.829 (blue) respectively. The
grey-dotted line shows the FBGs’ cut with the lowest mass of about $10^{12.8}M_{\odot}/h$.
}
\end{figure}

Fig.\ref{Fig:pdf_obs} and Fig.\ref{Fig:pdf_sim} plot the absolute magnitude and mass function of galaxies and (sub)halos from redshift 0.18 to 0.28 in observation and simulation (without any measurement error) respectively. The grey-dotted lines in the two plots represent the cuts on the FBGs luminosity and (sub)halos mass respectively,  
from which we could tell that the samples we choose as FBGs and (sub)halos are both complete.



\subsection{Low Resolution Simulations With Different $\sigma_8$}
\label{simu_sigma8}

One of the main purpose of this work is to constrain the cosmological parameter $\sigma_8$ (for simplicity, we fix all the other cosmological parameters). This requires us to have theoretical predictions for the three statistics with different $\sigma_8$. We run GADGET \cite{2001NewA....6...79S,2005MNRAS.364.1105S} to perform a set of N-body simulations with six different values of $\sigma_8$ (0.73, 0.76, 0.79, 0.829, 0.86, 0.89). Note that the value of 0.829 is from CG. Every simulation involves $512^3$ dark matter particles in a $600^3 (h^{-1}Mpc)^3$ periodic box, and evolves from redshift $z_{ini}=72$ to the present day z=0. The particle mass is $1.4 \times 10^{11}h^{-1}M_{\odot}$, which is eight times the particle mass in CG. These simulations have the same random seed and the same initial conditions to minimize the influence from the cosmic variance. We use the FoF and HBT algorithm to identify halos and subhalos.

Note that to generate mocks in these small simulations, we only use one single snapshot of redshift 0.23, and choose our mocks in this cubic instead of 
building lightcones. 
 
Note that we do not include masks in our small simulations, as we find that the statistics of 2PCF, GGL and LDP do not change much as long as the lowest mass of FBGs in these simulations is fixed at the value of CG. More details regarding this point are given in Appendix B.

By combining the measurements of 2PCF, GGL and LDP lensing from CG and the six small simulations,
we can derive the statistics at different values of $\sigma_8$ using the following formula:

\begin{equation}
    \hat{\xi}(\sigma_8)=\hat{\xi}^S(\sigma_8)-\hat{\xi}^S(\sigma_8^{*})+\hat{\xi}^{CG}(\sigma_8^{*})
\end{equation}
in which $\hat{\xi}$ stands for any one of the three statistical quantities we are interested. The upper-index $S$ refers to the results from the small simulations, while the upper-index $CG$ stands for those from CG.  
This procedure combines the high and low resolution simulations, which provide the accurate measurement and the estimation of different cosmology respectively. It has been practiced in \cite{2023ApJ...953...98D}.
Note that $\sigma_8^{*}$=0.829, which is the value of the fiducial model used in CG. The other two parameters ($\sigma_{\lg M}$ and $\sigma_z$) in the small simulations are kept to be the same as those of CG. To identify the FBGs in the small simulations, we follow the same SHAM procedures shown in Fig.\ref{Fig:flow chart} to add mass and redshift scatters\footnote{Although we only use one snapshot of redshift 0.23, we also move the (sub)halos along LoS to mimic the redshift scatter in observation.}.

\section{Main Results}
\label{result} 

\subsection{Direct Results from DECaLS and CG}

For FBGs in DECaLS, we choose the absolute magnitude cut to be -21.5 in r-band, the redshift range is from 0.18 to 0.28, we also do some cuts for the background source: 1) the signal-to-noise-ratio defined in Fourier space \cite{Li_2021} is required to be larger than 4 for avoiding selection/detection biases at the faint end in shear measurement; 2) the photometric redshift of the background galaxy should be larger than 0.43 (= 0.2 + the median FBGs) for avoiding significant overlaps between the lens and the source populations.  

\begin{figure*}[ht]
    \includegraphics[width=0.96\textwidth]{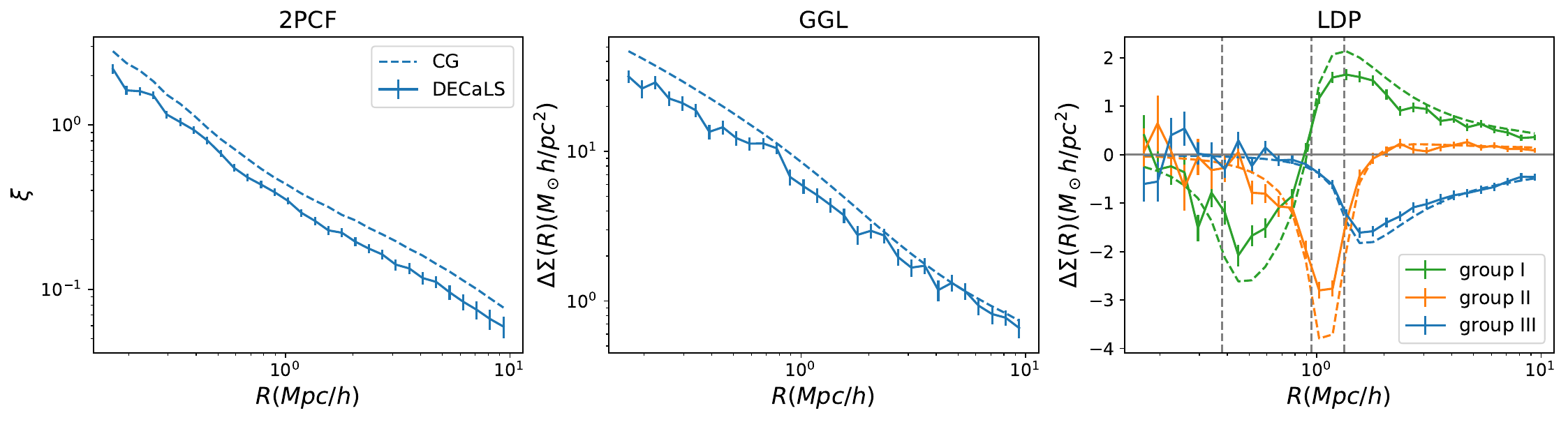}
    \caption{\label{Fig:pair+gg+ldp_origin}
   The solid curves with error bars are the results of 2PCF, GGL and LDP lensing from DECaLS. The dashed lines are from CG without any measurement errors. The three groups of the LDPs ($2^\prime\le R_s<5^\prime$, $5^\prime\le R_s<7^\prime$ and $R_s\ge7^\prime$) are labeled as I, II, III respectively. The grey vertical dotted lines of the third panel represent $2^\prime, 5^\prime, 7^\prime$ respectively.   
}
\end{figure*}

The results of 2PCF, GGL and LDP lensing are shown in Fig.\ref{Fig:pair+gg+ldp_origin}. The lines with errorbars represent the observational results. The errorbars are estimated using 200 jackknife sub-samples\cite{1982jbor.book.....E}. For the three groups of LDP lensing, we plot them with green, orange, and blue colors respectively. The different choices of $R_s$ ($2^{\prime}$, $5^{\prime}$ and $7^{\prime}$) are shown with the grey dashed vertical lines. The results of the CG simulation are shown as the dotted lines, without adding redshift or halo mass uncertainties. Note that although our statistics are all measured in terms of the angular separation, we always plot the results using the comoving distance instead, which is calculated assuming that all the FBGs are at redshift 0.23. The distance range is from about 0.1 $Mpc/h$ to $10$ $Mpc/h$, equally divided into 30 logarithmic bins.

\begin{figure*}[ht]
    \includegraphics[width=0.96\textwidth]{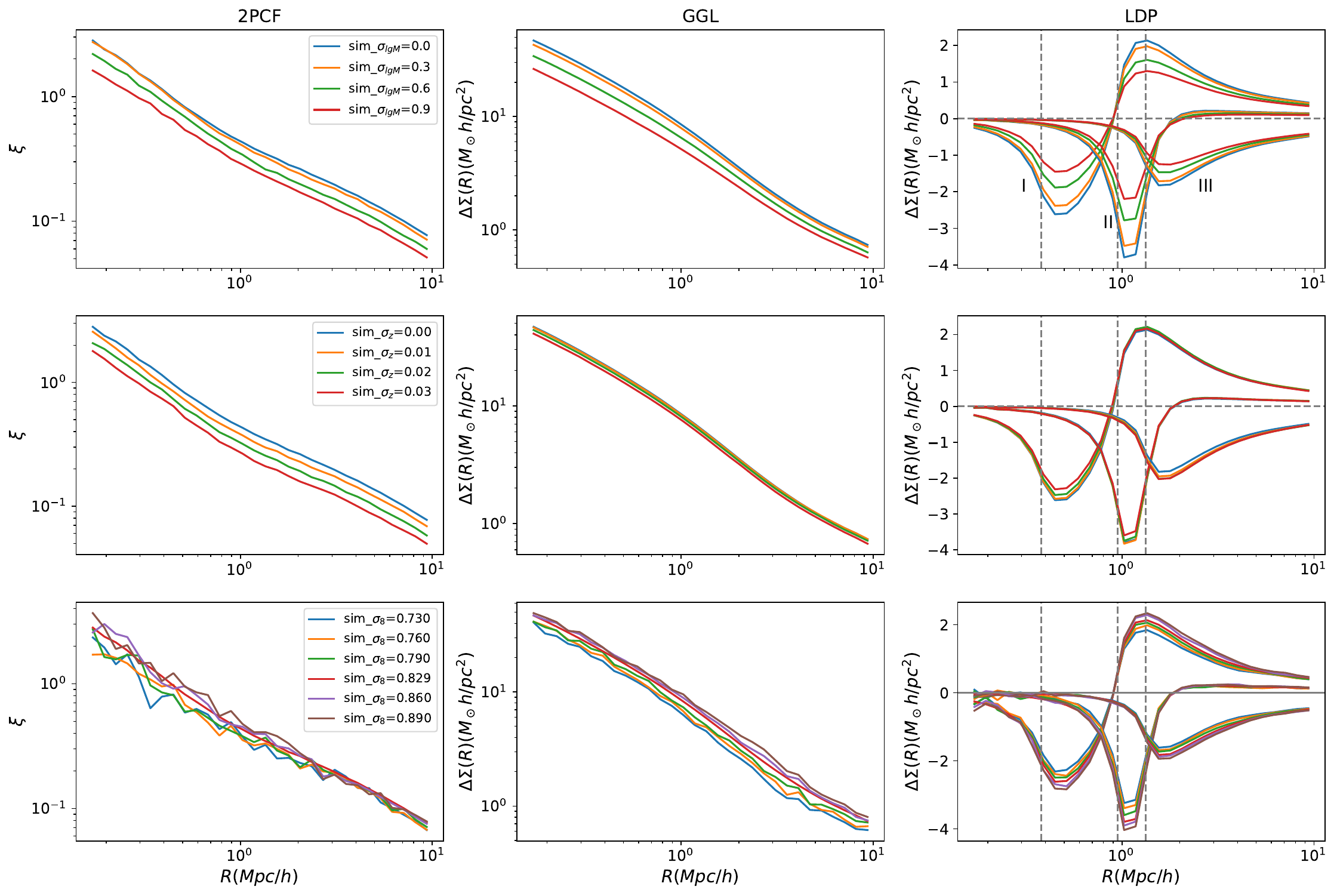}
    \caption{\label{Fig:pair+gg+ldp_M-21.5}
    The results of 2PCF, GGL and LDP lensing of simulation shown in the first, second and third column respectively. In each column, we show how the statistics change with different values of $\sigma_{\lg M}$, $\sigma_z$ and $\sigma_8$ in the upper, middle and lower panels respectively. Different colors refer to different measurement errors or $\sigma_8$ as labeled in the plots. The LDPs contain three groups labeled as I, II and III.   
}
\label{mz_dep}
\end{figure*}

The results show an obvious discrepancy between observation and simulation. The signals in simulation are generally larger than the observational ones by about $\sim 25$\%. We think this is mainly caused by the influence of measurement errors in observation or different cosmology, such as the photo-z error and the halo mass scatter in SHAM or the value of $\sigma_8$.
We try to mitigate the difference by adding different measurement errors in simulations or use different value of $\sigma_8$ to study the behavior of the three statistics under various circumstances.

\subsection{Impacts of Measurement Errors}

To get an idea about how the measurement errors affect the predictions on the statistics of our interest, we show in Fig.\ref{Fig:pair+gg+ldp_M-21.5} the dependence of 2PCF, GGL and LDP lensing on different assumptions about $\sigma_{\lg M}$ and $\sigma_z$ in the simulation. In the upper panels of Fig.\ref{Fig:pair+gg+ldp_M-21.5}, we use curves of different colors to represent results with different mass uncertainties. As the mass uncertainty increases, more low mass (sub)halos get involved in the SHAM, leading to weaker correlation between the (sub)halos and the density field, and therefore smaller signals of 2PCF and GGL. The mass uncertainties also makes the selected (sub)halos less representative of high density regions, causing weaker LDP lensing signals as well. 

In the middle panels of Fig.\ref{Fig:pair+gg+ldp_M-21.5}, we study the impact of the redshift uncertainty (in the absence of the mass uncertainty) on the three statistics respectively. In this case, the SHAM procedures replace some of the (sub)halos in the given redshift range with those massive ones from the neighboring redshift bins. We find that the 2PCF is decreased as the redshift scatter goes up. This is because the average 3D distances between the newly selected halos are increased. In contrast, the GGL signals only have mild changes, due to the fact that the halo mass function does not evolve rapidly with redshift, and the average mass of the selected (sub)halos is therefore not significantly changed. 

The behavior of the LDP lensing is somewhat nontrivial. In the inner regions, the LDP signal becomes modestly weaker as the redshift scatter increases, similar to the case of GGL, and for similar reasons. However, in the outer regions, the trend turns out to be the opposite. The main reason we believe is that the newly selected (sub)halos, although may seem to be more isolated than those excluded ones (due to redshift uncertainties) in the given redshift range, they still correctly identify the high density regions along the line of sight, therefore can help increase the significance of the LDP lensing signal on large scales.


\subsection{Dependence on $\sigma_8$}

The dependence of the three statistics on the value of $\sigma_8$ is shown in the lower panels of Fig.\ref{Fig:pair+gg+ldp_M-21.5}. Note that for this part of our study, we need small simulations of different $\sigma_8$, as mentioned in \S\ref{simu_sigma8}. As expected, the amplitudes of GGL and LDP lensing increase monotonically with $\sigma_8$. 

The change of 2PCF, on the other hand, is not obvious with the change of $\sigma_8$, especially for scales larger than 1 $Mpc/h$.
To understand this, for simplicity, let us consider only the linear region, in which the 2PCF can be written as: 
\begin{equation}
    \xi_{hh}=b_{h}^2\xi_{mm}\propto b_{h}^2\sigma_8^2,
\end{equation}
where $\xi_{mm}$ is the matter correlation, $\xi_{hh}$ is the halo-halo correlation, $b_{h}$ represents the average bias of the (sub)halos representing the FBGs. The halo bias can be estimated using the extended Press-Schechter formalism \cite{1996MNRAS.282..347M} as:
\begin{equation}
    b=1+\frac{\nu^2-1}{\delta_c},
\end{equation}
where $\nu=\delta_c/\sigma(M)$ is the parameter accounting for the so-called "peak height", $\delta_c=1.686$ is the linear overdensity threshold for halo collapse, and $\sigma\propto\sigma_8$ is the linear variance of matter. According to the formula, we know that for a universe with smaller $\sigma_8$, each given halo mass corresponds to a higher value of $\nu$, therefore a higher halo bias. It turns out that this effect largely counteracts the change of the mass power spectrum, making the 2PCF changes very mildly with $\sigma_8$. Similar findings have been reported in \cite{2012ApJ...745...16T} (Fig.3), although they applied the Halo Occupation Distribution (HOD) model instead of SHAM in their galaxy model. 
Using the halo bias formula proposed in \cite{2010ApJ...724..878T}, we indeed find that the average bias of the FBGs changes from 1.67 to 1.36 when $\sigma_8$ changing from 0.73 to 0.89, therefore the change of $\xi_{hh}$ is negligible ($\sim$0.5$\%$ from $\sigma_8$=0.73 to 0.89), as shown in the lower left panel of Fig.\ref{Fig:pair+gg+ldp_M-21.5}. The sensitivity of 2PCF to the value of $\sigma_8$ is very low. For this reason, we set the 2PCF to be independent of $\sigma_8$ in the rest of our analysis for simplicity, and we choose the default value of $\sigma_8^{*}$=0.829. In other words, we do not account for the constraining power of 2PCF on $\sigma_8$.

\subsection{Likelihood Analysis}
\begin{figure*}
    \centering
    \includegraphics[width=0.8\linewidth]{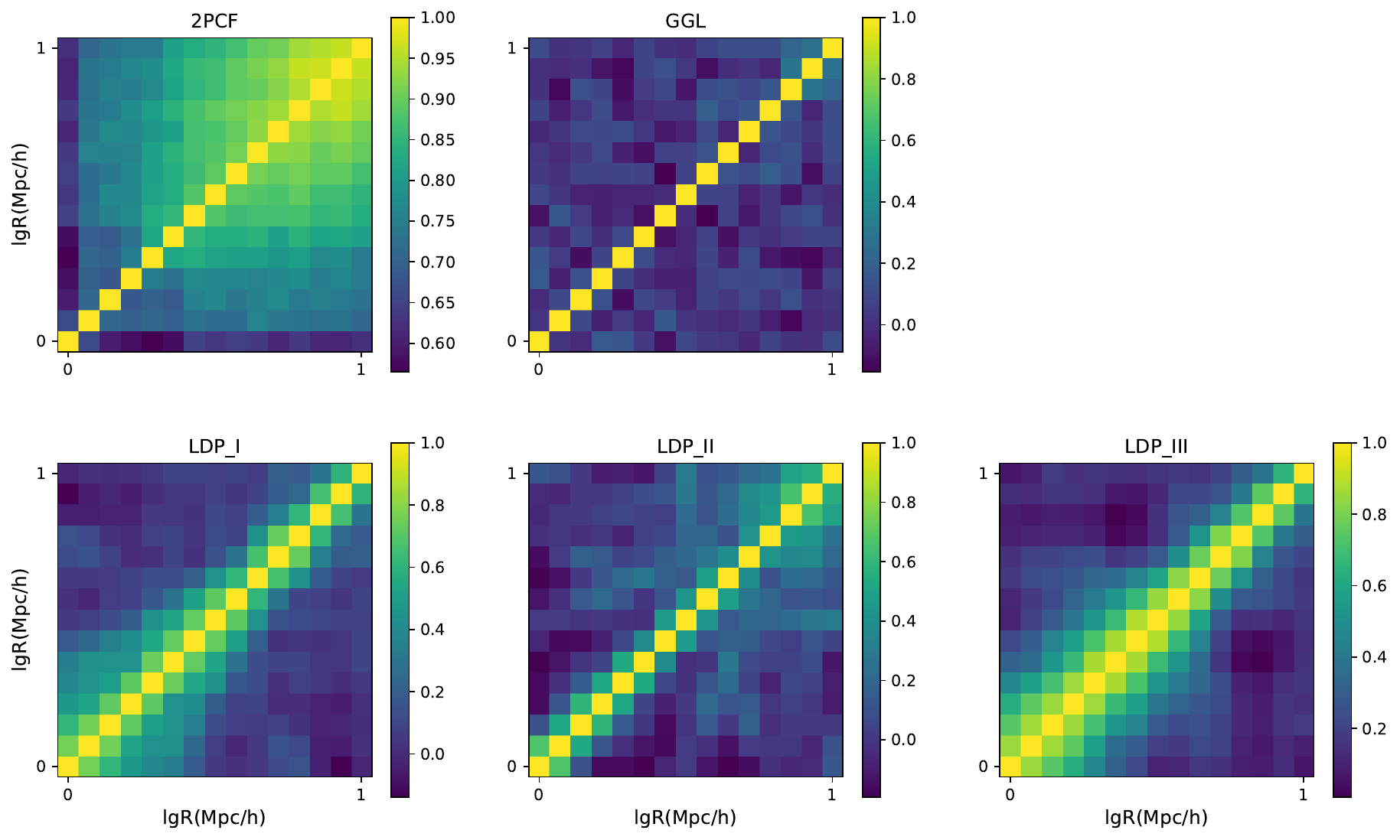}
    \caption{The normalized covariance matrix for 2PCF, GGL and the three groups of LDP in DECaLS used in the analysis, the axes of each plot are the logarithmic distribution of 15 bins from 1 to 10$Mpc/h$.}
    \label{fig:cm_obs}
\end{figure*}

We use the Bayesian analysis with the likelihood function defined as:
\begin{equation}
    \ln \boldsymbol{L(D |\Theta)=-\frac{1}{2}((D-M)^T C^{-1}(D-M))},
\end{equation}
where $\Theta$ represents the parameters we want to infer, $\boldsymbol{D}$ is the observational data measured from DECaLS, $\boldsymbol{M}$ is the simulation results, and $\boldsymbol{C}$ is the covariance matrix derived from Jackknife in observation. 
We use the K-means clustering algorithms \cite{2003PatRe..36..451L} to divide the samples (FBGs) 
into 200 sub-samples. For the LDPs, we use the FBGs to define the center of each subsample, and label each LDP according to its closest center. 
The covariance matrix of the observation data for each statistic used in the analysis is shown in Fig.\ref{fig:cm_obs}. Besides these individual covariance matrices, we also include the cross-correlation between each statistics in our analysis. 

Considering the error propagation that the error of the covariance matrix is transformed into the parameter fitting procedure \cite{2014MNRAS.439.2531P}, the covariance matrix should be rescaled to
\begin{equation}
    \tilde{C}=\frac{1+B(b-p)}{1+A+B(p+1)} C
\end{equation}
where p=3 is the number of parameters, b=15N is the number of bins (N is the number of statistics applied in the MCMC fitting), and
\begin{eqnarray}
&&A=\frac{2}{(n-b-1)(n-b-4)}\\ \nonumber
&&B=\frac{n-b-2}{(n-b-1)(n-b-4)}
\end{eqnarray}
with n=200, the number of jackknife sub-samples. One also needs to consider the noise in the original covariance matrix $C_{\ast}$\cite{2007A&A...464..399H}. The inverse of $C$ should be calculated as:
\begin{equation}
    C^{-1}= \frac{n-b-2}{n-1}C_{\ast}^{-1}.
\end{equation}

Based on Bayes analysis, the posterior distribution should be like,
\begin{equation}
    \boldsymbol{P(\Theta |D) \propto L(D|\Theta)p(\Theta) }
\end{equation}
where $\boldsymbol{p(\Theta)}$ is the prior distribution. We choose the range of $\sigma_{\lg M}$ and $\sigma_z$ to be (0.0,1.2) and (0.00,0.03)
, and take the samples with intervals 0.3 and 0.01 
respectively. 
We sample the two measurement errors along with the six values of $\sigma_8$ to make a theoretical table, according to the procedures shown in \S\ref{simulation}, and interpolate within this table using the package from Scipy\footnote{\url{https://scipy.org/}}. Then we use the emcee package\footnote{\url{https://emcee.readthedocs.io/en/stable}}\cite{2013PASP..125..306F} to perform the Markov Chain Monte Carlo sampling for $\sigma_{\lg M}$, $\sigma_z$ and $\sigma_8$. 
We only use the data points larger than 1$Mpc/h$ for avoiding the complex nonlinear evolution in the inner regions. We run the ensemble sampler with 50 walkers of 5000 steps for each chain and discard the first 100 steps. 

\begin{figure*}
    \includegraphics[width=0.96\textwidth]{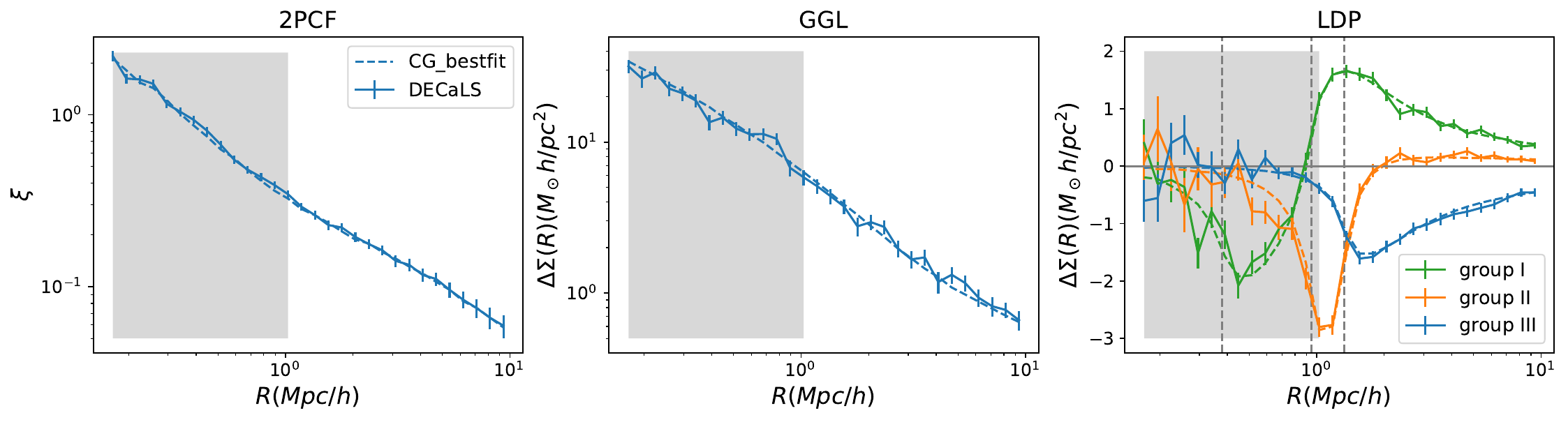}
    \caption{\label{Fig:fit}
    Similar as Fig.\ref{Fig:pair+gg+ldp_origin}, the dotted lines shows the best fit results using both the 2PCF, GGL and LDP lensing measurements, the blue lines with errorbar show the DECaLS results same as in Fig.\ref{Fig:pair+gg+ldp_origin}. The white area represent the data points we used to constrain $\sigma_{\lg M}$, $\sigma_z$ and $\sigma_8$. 
}
\end{figure*}

Fig.\ref{Fig:fit} shows the best-fit results in the MCMC calculations (dotted lines). The blue lines are the DECaLS results same as Fig.\ref{Fig:pair+gg+ldp_origin}. The grey shaded area represents the data points excluded for parameter estimation. All the three statistics show a good agreement with the observation results, which shows a feasible way to estimate the two errors $\sigma_{\lg M}$, $\sigma_z$ along with cosmological parameter $\sigma_8$.

\begin{figure*}
    \includegraphics[width=0.99\textwidth]{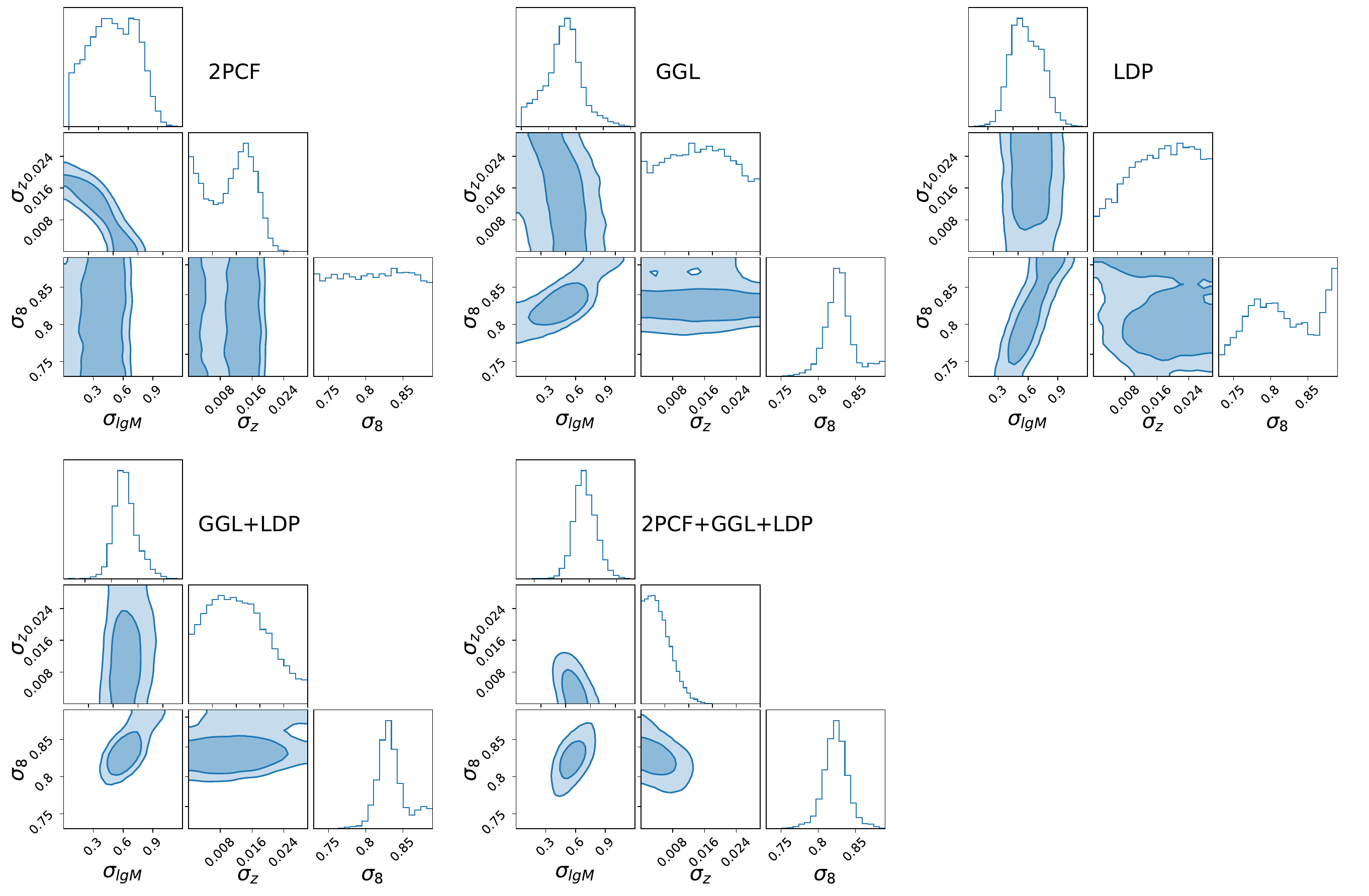}
    \caption{\label{Fig:contour}
    The 68\% and 95\% confidence level contour plots of $\sigma_{\lg M}$, $\sigma_z$ and $\sigma_8$ for 2PCF, GGL and LDP lensing, the combination of GGL and LDP, and the combination of all the three statistics.
}
\end{figure*}

In the upper panels of Fig.\ref{Fig:contour}, we show the 68\% and 95\% confidence contours for 2PCF, GGL, LDP lensing (combining group I, II and III together) respectively. One can see that the degeneracy directions between $\sigma_8$ and $\sigma_{\lg M}$ are significantly different for GGL and LDP lensing, therefore their combination can enhance the constraining power on the two parameters, as shown in the lower-left panel of Fig.\ref{Fig:contour}. On the other hand, in all cases, $\sigma_{\lg M}$ and $\sigma_z$ exhibit strong degeneracy, but with different directions. The combination of the three probes can therefore help break this degeneracy, as shown in the lower middle panel of Fig.\ref{Fig:contour}. Table \ref{tab:table1} shows the posteriors of each case.

The final constraint on $\sigma_8$ is $0.824^{+0.015}_{-0.015}$, which is consistent with the PLANCK result. But the results of $\sigma_{\lg M}$ and $\sigma_{z}$ are somewhat beyond our expectation. According to \cite{2008ApJ...676..248Y}, the average scatter of the lognormal luminosity distribution of central galaxies decreases from $\sim$0.375 mag at the massive end ($\lg [M_h/h^{-1}M_{\odot}] > \sim 13.5$) to $\sim$0.25 mag at the low-mass end ($\lg [M_h/h^{-1}M_{\odot}] \sim 12.0$). Considering the minimum mass of FBGs in simulation is from 12.74$h^{-1}M_{\odot}$ to 12.84$h^{-1}M_{\odot}$ (corresponding $\sigma_8$ from 0.73 to 0.89), our constraining result of $\sigma_{\lg M}$ is a little bit larger, meanwhile the result of $\sigma_{z}$ is a bit smaller than the mean photo-z error of the FBGs in DECaLS ($\sim$0.019). While both of the two measurement error results are reasonable considering the errorbar and the degeneracy of these two parameters.

Finally, the signal-to-noise (S/N) ratio of each statistics could be calculated as:
\begin{equation}
    \frac{S}{N}=[\sum_{i,j}\Delta D(\theta_i)C_{i,j}^{-1}\Delta D(\theta_j)]^{1/2}
\end{equation}
where D represents the three statistics in observation and the summation is over the radius bins of the fitting range. 
For the data we use for the parameter fitting ($>$ 1 $Mpc/h$), the S/N are 26.1, 25.6 and 31.1 for 2PCF, GGL and LDP lensing (combining three groups together) respectively. 
It shows the potential importance of LDP lensing in cosmological constraints.


\begin{table}
    \centering
    \renewcommand{\arraystretch}{1.2}
    \begin{tabular}{c|c|c|c}
    \hline statistical probes
    & $\sigma_{\lg M}$  & $\sigma_z$ & $\sigma_8$  \\ \hline
    Priors & [0.0,1.2] & [0.00,0.03] &[0.73,0.89] \\ \hline\hline
     2PCF    &$0.373^{+0.210}_{-0.214}$
     &$0.011^{+0.005}_{-0.009}$
     &-  \\ \hline
    GGL &$0.473^{+0.163}_{-0.212}$ & $0.015^{+0.009}_{-0.010}$ &$0.828^{+0.018}_{-0.017}$ \\ \hline
    LDP &$0.623^{+0.175}_{-0.140}$ & $0.017^{+0.008}_{-0.009}$ &$0.813^{+0.061}_{-0.044}$ \\ \hline
    GGL+LDP &$0.625^{+0.117}_{-0.092}$ & $0.012^{+0.008}_{-0.007}$ &$0.832^{+0.023}_{-0.014}$ \\ \hline
    2PCF+GGL+LDP &$0.565^{+0.086}_{-0.070}$ & $0.004^{+0.004}_{-0.003}$ &$0.824^{+0.015}_{-0.015}$ \\ \hline

    \end{tabular}
    \caption{\label{tab:table1} The priors of $\sigma_{\lg M}$, $\sigma_z$ and $\sigma_8$ and the posteriors for individual 2PCF, GGL, LDP, the combination of GGL and LDP, and the combination of all for DECaLS.}
    \label{result1}
\end{table}


\section{Method Validation with Simulation} 
\label{sec:method validation}
To test the reliability of our method, we use simulation as mocks to do the validation test.
Since we have nine lightcones in total as introduced in \S\ref{lightcone}, we use one of them to mimic the observation by adding mass and redshift errors. We take the average of the other eight lightcones to make our model predictions. We apply the same procedure to constrain $\sigma_{\lg M}$, $\sigma_z$ and $\sigma_8$ to test if we could reproduce their input values. 
As for the covariance matrix, we use the one from the real data directly (Fig.\ref{fig:cm_obs}), rather than calculating it from the mock. This is for including more observational effects, such as the shape noise of the source galaxies. 

\begin{figure}
    \centering
    \includegraphics[width=0.7\linewidth]{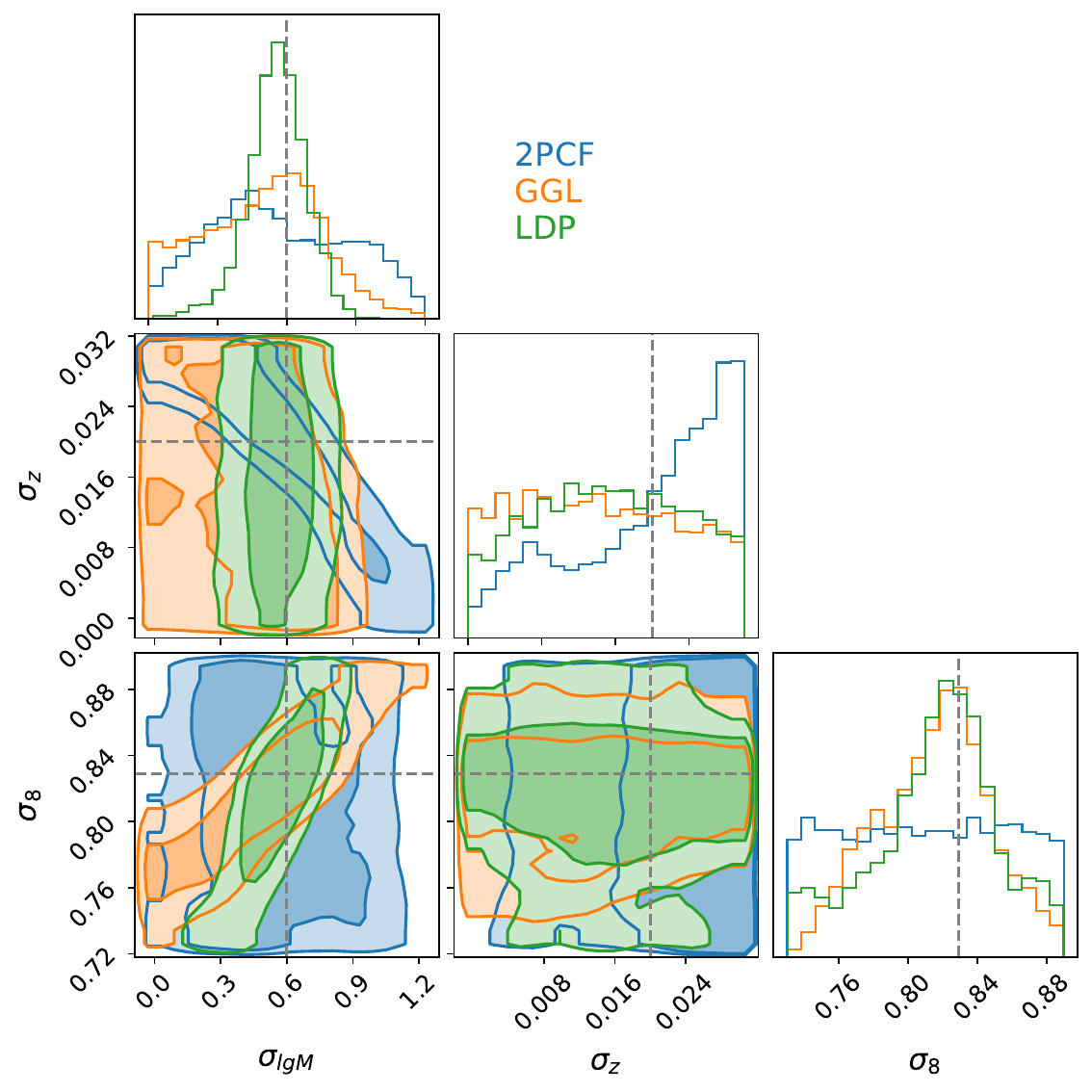}
    \caption{In the validation test, the posterior distributions of $\sigma_{\lg M}$, $\sigma_z$ and $\sigma_8$ for 2PCF, GGL and LDP lensing, plotted as the blue, orange and green contours respectively. The grey dotted lines represent the true values of each parameter, i.e., $\sigma_{\lg M}=0.6$, $\sigma_z=0.02$ and $\sigma_8=0.829$.}
    \label{fig:contour_slice2}
\end{figure}

\begin{figure}
    \centering
    \includegraphics[width=0.7\linewidth]{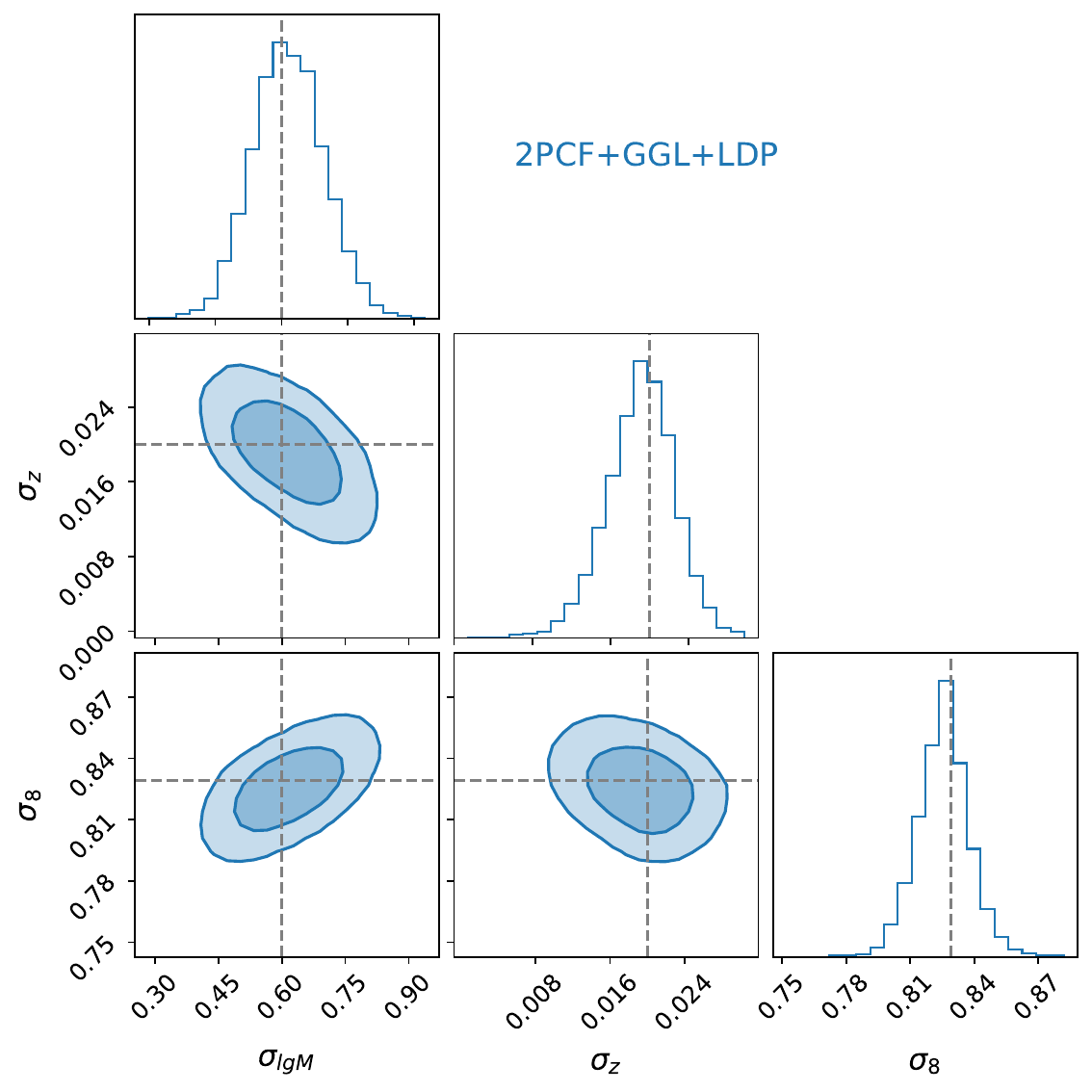}
    \caption{Posterior distributions for the combination of 2PCF, GGL and LDP lensing in the validation test. The grey dotted lines represents the true value of each parameter, i.e., $\sigma_{\lg M}=0.6$, $\sigma_z=0.02$ and $\sigma_8=0.829$.}
    \label{fig:contour_slice9}
\end{figure}

We choose $\sigma_{\lg M}=0.6$, $\sigma_z=0.02$, and $\sigma_8=0.829$ as the input values of the parameters.
The constraining contours are shown in Fig.\ref{fig:contour_slice2} for 2PCF, GGL and LDP lensing (group I, II, III) individually, and in Fig.\ref{fig:contour_slice9} for their combination.
The numerical posterior results are shown in Table \ref{tab:table2}. We see that the degeneracy directions of the three parameters $\sigma_{\lg M}, \sigma_z, \sigma_8$ for 2PCF, GGL and LDP are similar to those shown in Fig.\ref{Fig:contour}. 
The combined results in Fig.\ref{fig:contour_slice9} does demonstrate the robustness of our method in constraining the parameters.

\begin{table}
    \centering
    \renewcommand{\arraystretch}{1.2}
    \begin{tabular}{c|c|c|c}
    \hline statistical probes
    & $\sigma_{\lg M}$  & $\sigma_z$ & $\sigma_8$  \\ \hline
    Priors & [0.0,1.2] & [0.00,0.03] &[0.73,0.89] \\ \hline
    Fiducial value & 0.6 & 0.02 & 0.829 \\ \hline\hline
     2PCF    &$0.532^{+0.388}_{-0.280}$
     &$0.022^{+0.006}_{-0.013}$
     &-  \\ \hline
    GGL &$0.509^{+0.226}_{-0.316}$ & $0.014^{+0.010}_{-0.009}$ &$0.818^{+0.029}_{-0.039}$ \\ \hline
    LDP &$0.561^{+0.116}_{-0.119}$ & $0.015^{+0.009}_{-0.009}$ &$0.820^{+0.029}_{-0.042}$ \\ \hline
    2PCF+GGL+LDP &$0.615^{+0.081}_{-0.078}$ & $0.019^{+0.003}_{-0.004}$ &$0.826^{+0.011}_{-0.012}$ \\ \hline

    \end{tabular}
    \caption{\label{tab:table2}The priors and the fiducial values of $\sigma_{\lg M}$, $\sigma_z$ and $\sigma_8$, and the posteriors for individual 2PCF, GGL, LDP and the combination in the validation test.}
    \label{result_tab}
\end{table}

\section{Conclusion and Discussion}

Large scale galaxy imaging surveys offer excellent opportunities for exploring the details of structure formation as well as testing the consistency of the underlying cosmological models. In this work, we use the DECaLS DR9 galaxy catalog and the shear catalog based on the DECaLS DR8 imaging data to study three types of statistics: the two-point correlation function (2PCF) of the foreground bright galaxies (FBGs) in projection; the stacked galaxy-galaxy lensing (GGL) signals of the FBGs; the stacked lensing signals around a set of low-density positions (LDPs). Our FBGs are defined simply by applying a cut on the r-band absolute magnitude of the galaxies within a narrow photo-z range. The LDPs are defined by excluding the neighborhood of the FBGs within a specified radius ($R_s$). It was first proposed by \cite{2019ApJ...874....7D}. Comparing to the traditional ways of defining the low density regions/voids, LDPs only need photo-z information and generally yield a higher significance for the stacked lensing signal. As the three probes provide highly complementary information of large scale structure, we expect that the data set can help us address questions regarding the consistency of the $\Lambda$CDM model, in particular the 'lensing-is-low' problem.

Our theoretical predictions are from N-body simulations. 
We apply one of the high resolution CosmicGrowth simulations, 
and use SHAM to link the (sub)halos to galaxies. Without introducing any mass scatter in SHAM, nor any redshift uncertainties, we show in Fig.\ref{Fig:pair+gg+ldp_origin} a direct comparison of the three statistics between the observation and simulation. The results from CG are generally stronger than those of DECaLS, and the difference can reach up to $\sim25\%$. We believe the main drivers of this discrepancy include the mass scatter in the halo-galaxy correspondence and the redshift uncertainties of the photo-z catalog. This is demonstrated in the upper two rows of panels of Fig.\ref{mz_dep}. Another important factor we consider is the $\sigma_8$ parameter. In the last row of Fig.\ref{mz_dep}, we also show the dependence of the three statistics on the value of $\sigma_8$, calculated with a number of smaller size simulations. 

Fig.\ref{Fig:contour} further demonstrates quite different dependencies of 2PCF, GGL and LDP on $\sigma_{\lg M}$, $\sigma_z$ and $\sigma_8$: 2PCF strongly depends on $\sigma_{\lg M}$ and $\sigma_z$ in a highly degenerate way, but not so much on $\sigma_8$; both GGL and LDP lensing show weak dependence on $\sigma_z$, but are sensitive to $\sigma_{\lg M}$ and $\sigma_8$ with different degeneracy direction. 
The combination of these statistics exhibits a strong ability of constraining all three parameters, as shown in the last panel of Fig.\ref{Fig:contour}. The numerical results are given in Table \ref{result1}. The best-fit curves also matches the observational results of DECaLS very well, as shown in Fig.\ref{Fig:fit}. Our constraint on $\sigma_8$ ($0.824^{+0.015}_{-0.015}$) is consistent with PLANCK. Nevertheless, we are aware that the best fit value of $\sigma_{\lg M}$ 
 (=$0.565^{+0.086}_{-0.070}$) is
somewhat larger than expected, and also the value of $\sigma_z$ (=$0.004^{+0.004}_{-0.003}$) is smaller than the official result of DECaLS ($\sim$0.019).  This may be caused by the simple assumption of mass and photo-z uncertainties in \S\ref{4.4}, e.g., we did not consider the mass dependence of $\sigma_{\lg M}$, nor the redshift
dependence of $\sigma_z$ or any sort of catastrophic outliers in
the photo-z error distribution. The current consistency on $\sigma_8$ with Planck could be possibly due to the insufficient modeling of the systematic effects. Nevertheless, we have demonstrated the importance of including the mass scatter and redshift uncertainty in the modeling of 2PCF, GGL and LDP lensing.

We test our method using mocks generated also from the same simulations. Instead of taking the average of the nine lightcones as the theoretical prediction, we randomly select one of them as the 'observational' case, and take the average of the other eight for theory. For the 'observational' lightcone, we add $\sigma_{\lg M}$=0.6 and $\sigma_z$=0.02 to the mock, then use the other eight lightcones to test the validation of our method. The constraints from the three statistics shown individually in Fig.\ref{fig:contour_slice2} are similar to those in Fig.\ref{Fig:contour}. The different parameter-degenerate directions imply a strong constraining power for the combination of the three probes, as  shown in Fig.\ref{fig:contour_slice9}. The best fit results of the three parameters are consistent with their input values.

\begin{figure*}
    \includegraphics[width=0.99\textwidth]{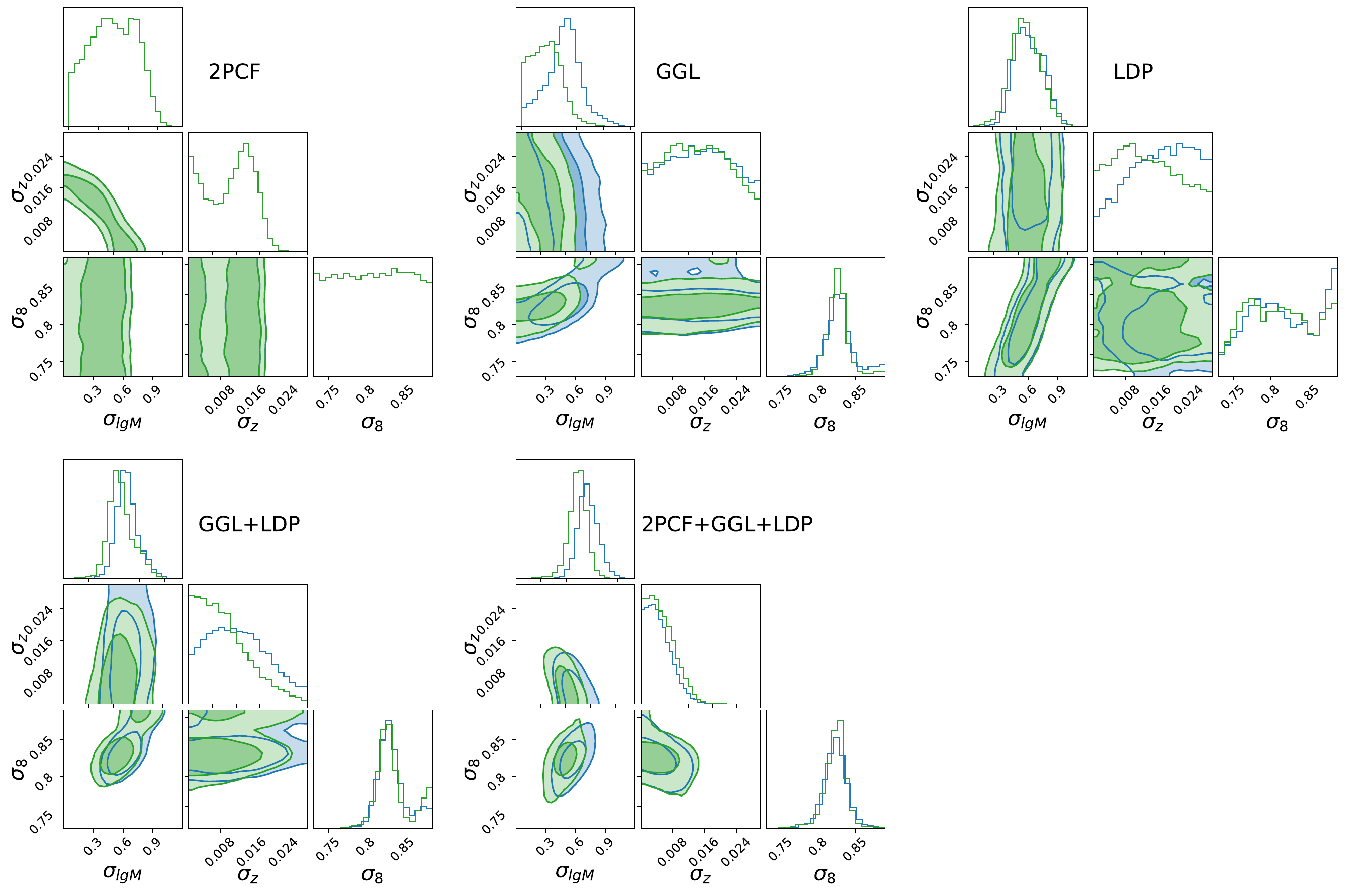}
    \caption{\label{Fig:contour_diffphotoz}
     The blue contour is the same as that in Fig.\ref{Fig:contour}. The green contour shows the results with the source redshift cut at 0.48,  which is 0.05 larger than the fiducial case of this work. The two results are consistent with each other.
}
\end{figure*}

Besides the validation of the method, for the stability of the parameter inference, we investigate the photometric redshift influence of the source galaxy, which is known as the dilution effect that some galaxies are misidentified as sources so that the weak lensing signal is diluted due to the fact that these galaxies are not truly sheared. So for testing the reliability in inferring cosmological parameter $\sigma_8$,  we apply a more restricted redshift selection criteria for the source galaxies: they should be larger than the median value of the lens redshift by 0.25 instead of 0.2, i.e. z$\geq$0.48. The results are shown as the green contours of Fig.\ref{Fig:contour_diffphotoz}. They are well consistent with the previous results.

Our work demonstrates that LDP lensing could enhance the constraining power in parameter inference together with the statistics of 2PCF and GGL. We also provide a way to estimate the possible measurement errors by mimic them in simulation. For simplicity, we have only considered the Gaussian distribution of $\sigma_{\lg M}$ and $\sigma_z$, and only one cosmological parameter $\sigma_8$. For future applications, we plan to carry out similar analysis with large simulations/emulators \citep {2019MNRAS.484.5509E,2021MNRAS.507.5869A,2025SCPMA..6889512C}, a larger cosmological parameter space, and more realistic assumption about the measurement errors.

\acknowledgments

We thank Xiaokai Chen, Hongyu Gao, Feihong He, Zhenjie Liu, Jiaqi Wang , Yihe Wang, Haojie Xu, and Ji Yao for helpful discussing. Thanks to the CosmicGrowth Simulation.
This work is supported by the National Key Basic Research and Development Program of China (2023YFA1607800, 2023YFA1607802), the NSFC grants (11621303, 11890691, 12073017), and the science research grants from China Manned Space Project (No. CMS-CSST-2021-A01).
F.D. acknowledges the financial support from the National Natural Science Foundation of China, grant No.12303003.
The computation resources of this work are provided by the Gravity Supercomputer at the Department of Astronomy and the $\pi$2.0 cluster of the Center for High Performance Computing, Shanghai Jiao Tong University.

\bibliographystyle{JHEP}
\bibliography{main.bib}

\appendix
\section{Selection Bias Correction}
\begin{figure*}
    \centering
    \includegraphics[width=0.9\linewidth]{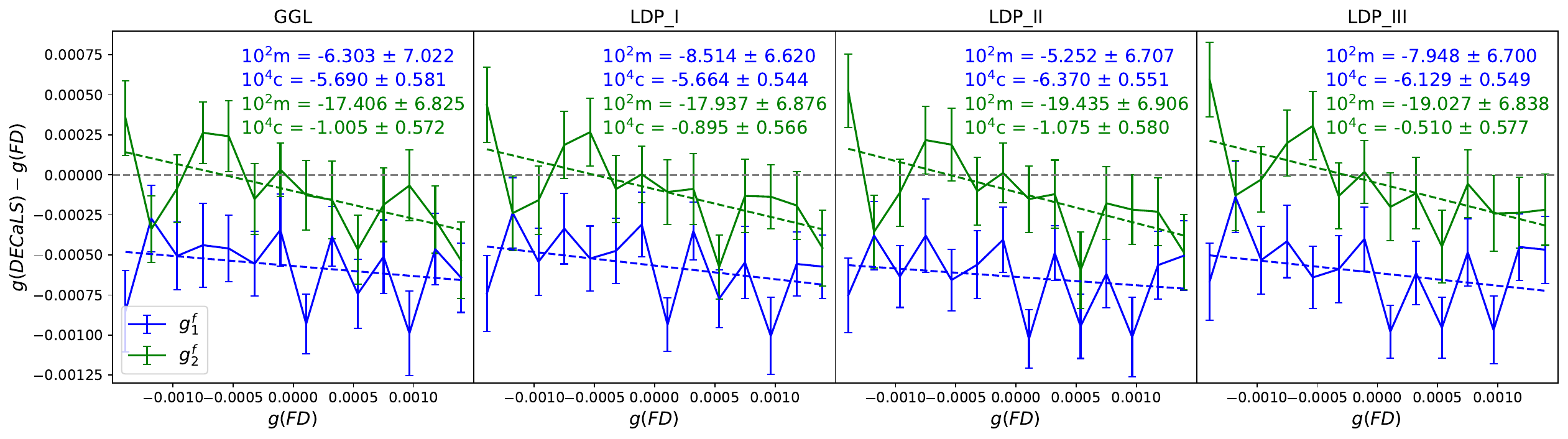}
    \caption{The difference between galaxy shear and field distortion for GGL and the three groups of LDPs respectively, the blue and green lines represent the shear component $g_1^f$ and $g_2^f$ respectively.}
    \label{fig:FD}
\end{figure*}

In a recent work, it has been found that a cut on the photo-z of the background galaxy can introduce additional selection biases \cite{2025JCAP...01..068S}. This occurs whenever a selection based on the photo-z of the source galaxy is applied. In this work, since we need to choose the galaxies that are on the background of the lenses, a photo-z cut is inevitable. Fortunately, the FQ shear catalog does retain the FD information for each set of shear estimators (on individual exposures), we are thus able to perform an onsite calibration of the shear biases using only the relevant subset of the catalog. 
We also consider the influence of $\Sigma_{c}$ for different lensing pairs, (i.e. during the stacking of the shear signal, for each pair, we count the number as $\Sigma_{c}$ instead of 1), the results are shown in Fig.\ref{fig:FD}.

We correct the shear estimators $G_1$ and $G_2$ with m and c according to Fig.\ref{fig:FD}:
\begin{eqnarray}
    &G_1=(G_1-c_1 \times(N+U)-c_2\times V)/(1+m_1)\\
    &G_2=(G_2-c_2 \times(N-U)-c_1\times V)/(1+m_2).
\end{eqnarray}

\section{Irrelevance of the Mask}
\begin{figure}
    \centering
    \includegraphics[width=0.96\linewidth]{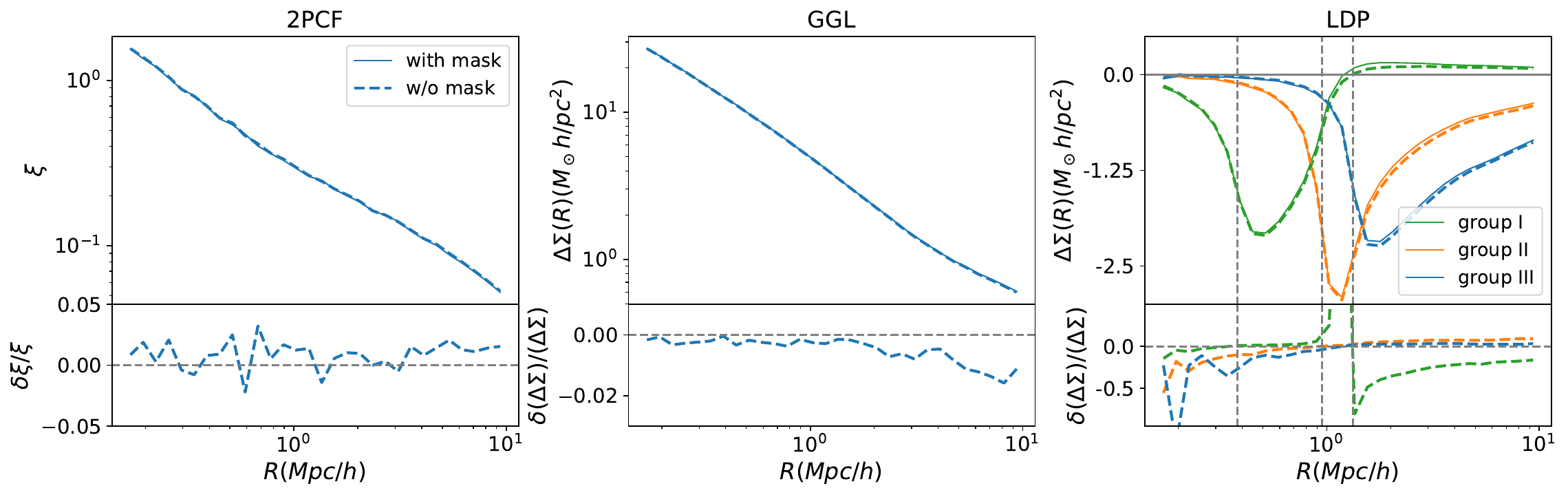}
    \caption{2PCF, GGL and LDP lensing in CG with (solid lines)/without (dashed lines) masks. The lowest mass of the FBGs is fixed at $10^{12.32}M_{\odot}/h$. The upper panels are the signals, the lower panels show the relative difference of each statistics.}
    \label{fig:mask}
\end{figure}

In measuring 2PCF, GGL and LDP lensing of the six low-resolution simulations, we do not include masks. We find that if the galaxy number density/the lowest mass of FBGs maintains the same, adding mask or not makes negligible difference. We test this using the original CG simulation without any measurement errors. The lowest mass of the FBGs is fixed at $10^{12.32}M_{\odot}/h$. The results are shown in Fig.\ref{fig:mask}. The solid curves show the results with the DECaLS masks, and the dashed ones are those without masks. As shown in the lower panels, all the three statistics show negligible relative difference (the difference between signals without/with mask divide by the signals with mask) according to the figure. The exception is the LDP lensing of group I at around 1 $Mpc/h$, which is caused by the smallness of the signal itself.   

\end{document}